\documentclass[prb,showpacs,twocolumn, amsmath,amssymb]{revtex4}

\usepackage{graphicx}
\usepackage{dcolumn}
\usepackage{bm}
\newdimen\figurewidth
\begin{document}
\preprint{APS/123-QED}

\title{Numerical study of transport through a single impurity\\
in a spinful Tomonaga-Luttinger liquid}

\author{Yuji Hamamoto}
\email{hamamoto@issp.u-tokyo.ac.jp}
\author{Ken-Ichiro Imura}%
\author{Takeo Kato}
\affiliation{%
Institute for Solid State Physics, The University of Tokyo, Kashiwa,
Chiba 277-8581
}%


\date{\today}

\begin{abstract}
The single impurity problem in a spinful Tomonaga-Luttinger liquid is studied numerically 
using path-integral Monte Carlo methods. 
The advantage of our approach is that the system allows for extensive analyses of 
charge and spin conductance in the non-perturbative regime. 
By closely examining the behavior of conductances at low temperatures, in the presence of 
a finite backward scattering barrier due to the impurity, we identified four distinct phases 
characterized by either perfect transmission or reflection of charge and spin channels. 
Our phase diagram for an intermediate scattering strength is consistent with the standard 
perturbative renormalization group (RG) analysis in the limit of weak and strong backward 
scattering, in the sense that all our phase boundaries interpolate the two limiting cases. 
Further investigations show, however, that precise location and form of our phase boundaries 
are not trivially explained by the standard RG analysis, e.g., 
some part of the phase diagram looks much similar to the weak backscattering limit, 
whereas some other part is clearly derived from the opposite limit. 
In order to give a more intuitive interpretation of such behaviors, we also 
reconsidered our impurity problem from the viewpoint
of a quantum Brownian motion picture.
\end{abstract}

\maketitle

\section{introduction}
Low-energy excitations in a one-dimensional electron system are
described by the \textit{so-called} Tomonaga-Luttinger liquid (TLL),
\cite{tomonaga}
which is characterized by power-low decay of various correlation
functions and the spin-charge separation.\cite{emery}
Recent progress in nano-fabrication techniques has enabled us to realize 
quasi-one-dimensional quantum structures, 
in which TLL behaviors have been experimentally observed in, e.g., 
fractional quantum Hall edges\cite{milliken}
and single-wall carbon nanotubes.
\cite{bockrath}
One way to highlight such peculiar behaviors of TLL is to introduce 
a single impurity, which dramatically influences the
transport of the system: in the low temperature limit,
a conducting channel turns either perfectly transmitting or insulating.
\cite{kane1,kane2,furusaki}\par

The impurity problem in a TLL is effectively described by bosonic fields
at the impurity,
which is equivalent to the problem of a quantum Brownian particle
moving in a periodic potential.
One of the powerful ways of treating such a complicated quantum
system is to numerically simulate the system using the path-integral Monte
Carlo (PIMC) method.
\cite{herrero}
Near the phase transition, however, a simulation
based on primitive local updates generally fails, and one calls
for a more effective update method optimized for the system. In
recent years, the efficiency of the PIMC
method has been remarkably improved by the extension of the Swendsen-Wang
cluster algorithm \cite{swendsen} to quantum systems with continuous degrees of freedom, and the
algorithm has been first applied to the phase transition
in a resistance-shunted Josephson junction system.\cite{werner1}\par

The impurity problem in both spinless and spinful TLLs has been originally
treated using perturbative renormalization group (RG) methods in
the weak- and strong-impurity limits.\cite{kane1,kane2,furusaki} 
Whether a conduction channel becomes perfectly transmitting or 
insulating at low temperatures is determined by the relevance of the corresponding
backscattering or tunneling process. 
In contrast to the spinless case, in which RG analyses in the weak- and
strong-backscattering limits seem to be smoothly connected,
the phase diagrams of the spinful case show clear
inconsistency in the two opposite limits, i.e., mismatch of RG flow,
suggesting the existence of an intermediate (unstable) fixed point.
Besides, the phase boundaries between conducting and insulating
phases for charge and spin are expected to shift continuously
as a function of the backscattering strength. 
Although one can see the essence of critical phenomena in this spinful
system using the standard perturbative RG approach, 
there is little information about the phase diagram for an impurity with
\textit{intermediate} strength. In this paper, we adopt the effective
PIMC simulations mentioned above as a non-perturbative approach to study critical
phenomena in the intermediate region of impurity strength.\par

Our spinful impurity problem can be also understood as the physics of
\textit{re-combination} of the electronic charge and spin, which are 
generally separated and propagate with different velocities in a bulk
TLL.
One can introduce the scattering effect of an impurity, say,
by annihilating one 
{\it physical}
electron from the right-going mode,
simultaneously creating another in the left-going mode.\cite{kane2,furusaki}
We mean by a 
{\it physical}
electron, an original electron composed of 
both charge and spin degrees of freedom.
If this scattering potential is relevant and grows stronger,
the charge and spin degrees of freedom become no longer independent
and their motion acquires some correlation.
At low temperatures, the effect of an impurity becomes dominant. As a result
the charge and spin tend to propagate almost together,
realizing a situation which we call spin-charge {\it re-combination}, 
unless the difference between their bulk velocities are too large.
Of course, if that difference is large enough,
the charge and spin could remain nearly independent, and the spin-charge separation
preserves. 
We will argue, based on our numerical results, 
whether or not the electronic charge and spin re-combine
depends complicatedly on the competition between the impurity strength and the
difference in the original charge and spin velocities.\par

This paper is organized as follows. 
In Sec. II we state our single impurity problem in a TLL --- the spinful
case with intermediate backward scattering strength. 
In Sec. III, we give details about the path-integral quantum Monte Carlo 
methods employed in this work.
In Sec. IV, we present our numerical results and the phase diagram
deduced from our data, and then we discuss them in comparison with
the known RG picture. Some further interpretation is also given in the
context of quantum Brownian motion.
Sec. V is devoted to the summary. 

\section{Statement of the problem}
Let us begin with introducing our model --- the single impurity problem
in a spinful 
Tomonaga-Luttinger liquid. Using the standard bosonization technique, we first 
formulate it in terms of two bosonic fields --- one for charge, and the other for spin.
Then, we rewrite it in a form suitable for numerical analyses.
We also briefly review what is known about our model in the standard perturbative 
RG picture. 
We end this section by addressing what we will attempt to uncover throughout this paper.

\subsection{The single impurity problem in a TLL --- the spinful case}
Low energy excitations of interacting one-dimensional electron
system with spin are density fluctuations of charge and spin, labeled
with subscripts $\rho$ and $\sigma$ respectively, and the
Hamiltonian is written as
\begin{gather}
H_0=\sum_{\nu=\rho,\sigma}\int\frac{{\rm
 d}x}{4\pi}\biggl[\frac{u_{\nu}}{K_{\nu}}\biggl(\frac{\partial\phi_{\nu}}{\partial
 x}\biggr)^2+u_{\nu}K_{\nu}\biggl(\frac{\partial\theta_{\nu}}{\partial x}\biggr)^2\biggr],
\end{gather}
where $\phi_{\nu}$ and $\theta_{\nu}$ are bosonic fields and the spatial
derivative of one field is the canonical momentum of the
other. $K_{\nu}<1$ for repulsive interaction
while $K_{\nu}>1$ for attractive, and $u_{\nu}$ is the
sound velocity of the density fluctuation. We now consider a single symmetric
impurity with finite reflection localized at the origin, and introduce
backward scattering of electrons by its barrier. Using the
bosonized representation of a fermionic operator $\psi_{rs}$ for an
electron moving in the direction $r=R$ or $L$ with spin $s=\uparrow$ or
$\downarrow$, the Hamiltonian corresponding to the lowest order
backscattering process is given by
\begin{align}
H_1&=V_0\sum_s\psi_{Ls}{}\!\!^{\dagger}(0)\,\psi_{Rs}(0)+{\rm H.c.}\nonumber\\
&=v\cos\phi_{\rho}(0)\cos\phi_{\sigma}(0).\label{bs}
\end{align}
Here $v$ is a parameter proportional to the scattering strength
$V_0$. Note that $\phi_{\nu}(0)/\pi$ denotes the number of charges (spins) for $\nu=\rho\,(\sigma)$ in the $x>0$ part of the system. Since the scattering term (\ref{bs}) influences only the fields at
the origin, we can integrate out the other fields away from the
barrier. If we write $\phi_{\nu}(\tau)\equiv\phi_{\nu}(x=0,\tau)$ in the
imaginary-time formalism, the effective action takes the form
\begin{align}
S&\equiv S_0+S_1,\label{act}\\
S_0&=\sum_{\nu}\sum_{\omega_n}\frac{|\omega_n|}{2\pi
 K_{\nu}\beta}|\tilde{\phi}_{\nu}(\omega_n)|^2,\label{act0}\\
S_1&=v\int_0^{\beta}{\rm
 d}\tau\cos\phi_{\rho}(\tau)\cos\phi_{\sigma}(\tau),
\end{align}
where $\tilde{\phi}_{\nu}(\omega_n)$ denotes the Fourier component of $\phi_{\nu}(\tau)$
and $\omega_n\equiv 2\pi n/\beta$ is the Matsubara frequency. $S_0$ is
known as the dissipative term in the Caldeira-Leggett model,
\cite{caldeira1} and can be
expressed by a form of long-range interactions in $\tau$ direction as
\begin{gather}
S_0=-\sum_{\nu}\frac{1}{2K_{\nu}\beta^2}\int_0^{\beta}{\rm
 d}\tau\int_0^{\beta}{\rm
 d}\tau'\frac{\phi_{\nu}(\tau)\phi_{\nu}(\tau')}{\sin^2[\frac{\pi}{\beta}(\tau-\tau')]},\label{long_range}
\end{gather}
which is used when we apply the cluster algorithm to the PIMC
simulation (see Sec. III).

\subsection{Consequences of the perturbative RG, and the three-dimensional 
RG phase diagram in the $\bm (\bm K_{\bm\rho}\bm,\bm
  K_{\bm\sigma}\bm,\bm v\bm)$-space}

\begin{figure}[h]
\begin{center}
\includegraphics[width=.5\figurewidth]{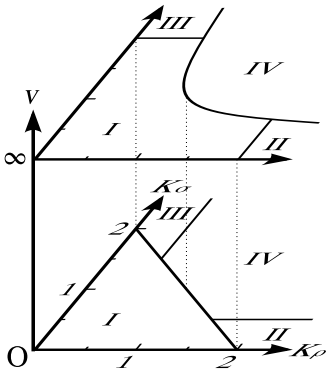}
\caption{\label{phase_rg}Three-dimensional phase diagram
in the $(K_{\rho},K_{\sigma},v)$-space. The phase boundaries are
analytically obtained in the following two limits:
for a weak impurity ($v\rightarrow 0$),
the phase boundaries are three straight lines $K_{\rho}+K_{\sigma}=2$,
$K_{\rho}=1/2$, and $K_{\sigma}=1/2$;
for a strong impurity ($v\rightarrow\infty$), the phase boundaries
are a hyperbola $K_{\rho}{}\!\!^{-1}+K_{\sigma}{}\!\!^{-1}=2$ and
two straight lines $K_{\rho}=2$ and $K_{\sigma}=2$. 
Three dotted lines represent those values of ($K_\rho,K_\sigma$) 
at which phase boundaries in the $v=0$ and $v=\infty$ plane coincide.
}
\end{center}
\end{figure}

In order to allow for a comparison of our numerical results
with the known analytic viewpoints, here we briefly review the
renormalization group (RG) picture presented in Refs. \onlinecite{kane2,furusaki}.
The standard perturbative RG analyses can be performed either for an
infinitesimal initial value of the scattering potential $v$
in the original model (\ref{act}) or in the opposite limit, i.e.,
for infinite backscatterings in the dual model of (\ref{act}).
They both give a phase diagram in the ($K_\rho, K_\sigma$)-plane
characterized by four different phases, which correspond to different 
transport behaviors of the system in the limit $T\rightarrow 0$:
(I) both charge and spin are insulating; 
(II) charge is conducting, while spin is insulating; 
(III) charge is insulating, while spin is conducting; 
(IV) both charge and spin are conducting. 
Are the phase boundaries between such four different phases
dependent on the initial values of $v$?
According to Refs. \onlinecite{kane2,furusaki},
the obtained phase diagram 
(Fig. \ref{phase_rg})
in the above two limits have, as expected, similar configurations, but
the phase boundaries are not located exactly at the same position in the
($K_{\rho},K_{\sigma}$)-plane. \par

Let us now ask a question, what happens if we start from an
intermediate value of the scattering potential $v$?
For such a value of $v$, one can {\it in principle} 
consider a phase diagram, analogous to the above two limiting cases,
i.e., probably with the same four distinct phases, 
but phase boundaries shifted from the two limiting cases. 
Since the bulk quantities $K_{\rho}$ and $K_{\sigma}$ are invariant 
under the RG transformation (because the barrier is localized at the origin),
we usually focus on a straight line connecting
($K_\rho, K_\sigma,0$) and ($K_\rho, K_\sigma,\infty$)
for a given set of $K_\rho$ and $K_\sigma$,
and examine how a scattering potential associated with a particular phase
scales in the RG transformation.
The results of Refs. \onlinecite{kane2,furusaki} show
that there exists a domain in the $(K_\rho,K_\sigma)$-plane,
in which this scattering potential is irrelevant in the limit of
$v\rightarrow 0$, whereas relevant in the opposite limit $v\rightarrow\infty$,
indicating the existence of a non-trivial fixed point
at an intermediate value of $v$ (this fixed point is shown to be unstable).
By performing the RG analyses one step further by considering
higher order perturbations, one finds non-linear RG equations,\cite{kane2}
suggesting a non-monotonous RG flow. 


To summarize, the standard RG approach,
with the help of duality transformation,
not only reveals the RG flow in the limit of weak-
and strong-backscattering barriers,
but, by extending the perturbative analysis one step further,
it also gives us some hints about how different tendencies of RG flow
in the two limits evolve and eventually merge
in the region of intermediate coupling.
On the other hand, there is little hope to obtain further information
on the RG flow in the whole parameter space,
most of which belongs to the so-called {\it non-perturbative} regime,
by simply elaborating such an analytical approach.
In this paper, we instead appeal to a numerical method, i.e.,
by performing a PIMC simulation for the effective action $S$
given in (\ref{act}), we study {\it directly} transport properties
of spinful TLL with an impurity of non-perturbative
backscattering potential barrier.

\section{simulation methods}
In this section, we illustrate our numerical simulations.
In order to eliminate critical slowing down at low temperature and carry out efficient
simulations of the paths $\phi_{\rho}(\tau)$ and
$\phi_{\sigma}(\tau)$,
we implement local updates in Fourier space and rejection-free global
updates following Refs. \onlinecite{werner1,werner2}.
Note that the single impurity problem in a spinful TLL is equivalent
to the overdamped limit of the Josephson junction system discussed
in Ref. \onlinecite{werner2}.
By discretizing the imaginary time into $N$ time steps,
we define $\phi_{\nu j}\equiv\phi_{\nu}(j\beta/N)\,(j=0,1,\cdots,N-1)$.
Then $S_0$ and $S_1$ can be rewritten as
\begin{align}
S_0&=\sum_{\nu=\rho,\sigma}\sum_{k=1}^{N/2}\frac{1}{2\sigma_{\nu
 k}{}^2}|\tilde{\phi}_{\nu k}|^2,\label{act_dissipation}\\
S_1&=v{\mit{\Delta}}\tau\sum_{j=0}^{N-1}\cos\phi_{\rho j}\cos\phi_{\sigma j},
\label{act_potential}\\
\sigma_{\nu k}{}^2&\equiv\left\{
\begin{array}{ll}
K_{\nu}N^2/4k&k=1,2,\cdots,N/2-1\\
K_{\nu}N&k=N/2
\end{array}
\right.\!\!,\label{variance}
\end{align}
where ${\mit{\Delta}}\tau=\beta/N$ and $\tilde{\phi}_{\nu
k}=\tilde{\phi}^{\ast}{}\!\!_{\nu,N-k}=\sum_j\phi_{\nu
j}\,e^{\frac{2\pi i}{N}jk}$.
In a local update for $k\ne 0$, a new value of $\tilde{\phi}_{\nu k}$ is randomly
generated from a normal distribution with the variance $\sigma_{\nu k}{}^2$
in (\ref{variance}) by means of the Box-M\"uller method.
This local update is accepted
with a probability
\begin{gather}
p={\rm min}\{1, e^{-{\mit\Delta}S_1}\},\label{prob_local}
\end{gather}
where ${\mit\Delta}S_1$ is the variation of the potential term (\ref{act_potential}).
For the $k=0$ component $\tilde{\phi}_{\nu 0}$, a new value is generated
from a uniform distribution ranged from $-\pi$ to $\pi$, and the local
update is accepted again with a probability (\ref{prob_local}).\par
\begin{figure}[t]
\begin{center}
\includegraphics[width=.8\figurewidth]{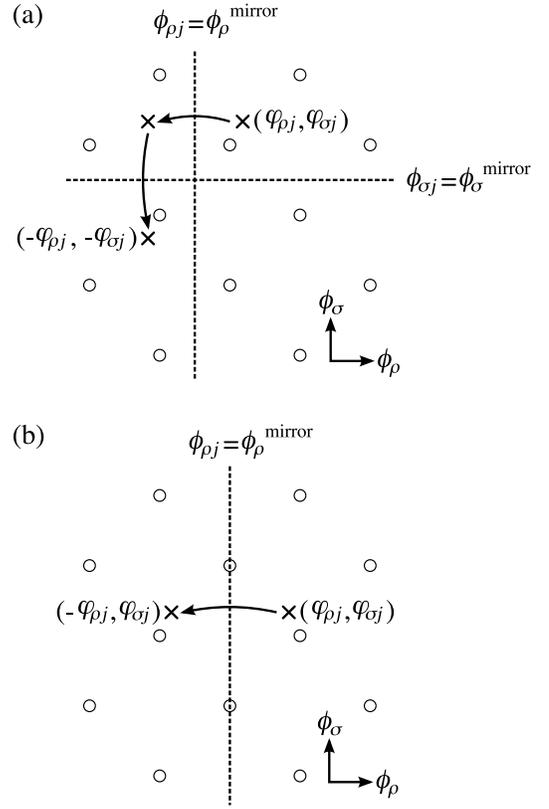}
\caption{\label{cluster_update}Cluster updates of field variables
$\phi_{\rho j}$ and $\phi_{\sigma j}$ in a $\phi_{\rho}$-$\phi_{\sigma}$
plane, where $\varphi_{\nu j}\equiv\phi_{\nu j}-\phi_{\nu}{}^{\rm
 mirror}$.
The empty circles represent the potential minima,
and the dashed lines represent the reference fields
$\phi_{\rho}{}^{\rm mirror}$ and $\phi_{\sigma}{}^{\rm mirror}$.
Only the field variables at the $j$-th time step in a cluster are shown.
(a) During one period of a double-field cluster update,
a point ($\varphi_{\rho j}, \varphi_{\sigma j}$) is subject to mirror
reflection twice, i.e.,
once with respect to $\phi_{\rho j}=\phi_{\rho}{}^{\rm mirror}$,
and subsequently to $\phi_{\sigma j}=\phi_{\sigma}{}^{\rm mirror}$.
(The whole process is
$(\varphi_{\rho j},\varphi_{\sigma j})\rightarrow(-\varphi_{\rho j},
\varphi_{\sigma j})\rightarrow(-\varphi_{\rho j},-\varphi_{\sigma j})$.)
(b) As for a charge-field cluster update,
$\varphi_{\nu j}$ is subject to mirror reflection with respect to
$\phi_{\rho}=\phi_{\rho}{}^{\rm mirror}$, i.e.,
$(\varphi_{\rho j},\varphi_{\rho j})\rightarrow(-\varphi_{\rho j},\varphi_{\rho j})$.
Similary, $\varphi_{\sigma j}$ in a spin-field cluster is updated as
$(\varphi_{\rho j},\varphi_{\sigma j})\rightarrow(\varphi_{\rho j},-\varphi_{\sigma j})$.
}
\end{center}
\end{figure}

A global update scheme
should be designed so that optimized paths for a given potential are
efficiently generated. In the case of the double cosine potential (\ref{bs}),
an optimized path near the phase transition typically spends most of the
time in potential minima,
and also has some kink structures connecting adjacent potential minima,
i.e., $(\phi_{\rho},\phi_{\sigma})=(n_{\rho}\pi,n_{\sigma}\pi)$ with
integers $n_{\rho}$ and $n_{\sigma}$ such that
$n_{\rho}+n_{\sigma}={\rm odd}$ for $v>0$.
In order to generate such kinks, we apply the Swendsen-Wang
algorithm\cite{swendsen} to update of the continuous field variable $\phi_{\nu
j}$ following Ref. \onlinecite{werner1}. To this end, we introduce a relative field
variable $\varphi_{\nu j}\equiv\phi_{\nu
j}-\phi_{\nu}{}^{\rm mirror}$ as shown in Fig. \ref{cluster_update},
where the reference $\phi_{\nu}{}^{\rm mirror}$ is appropriately chosen as described below.
If we regard the sign of the relative field $s_{\nu j}\equiv\varphi_{\nu
j}/|\varphi_{\nu j}|$ as a spin variable, the dissipative term in
(\ref{long_range}) can be represented as a kind of one-dimensional long-range Ising model
\begin{align}
S_0&=-\sum_{\nu}\sum_{j<j'}\kappa_{\nu jj'}s_{\nu j}s_{\nu j'},\label{act_spin}\\
\kappa_{\nu jj'}&\equiv\frac{|\varphi_{\nu j}||\varphi_{\nu j'}|}{K_{\nu}N^2}\frac{1}{\sin^2[\frac{\pi}{N}(j-j')]},
\end{align}
where each site labeled by $j$ corresponds to each time step $\tau_j\equiv
j{\mit\Delta}\tau$,
and has two spin variables $s_{\rho j}$
and $s_{\sigma j}$.
Due to the finite bandwidth cutoff, we should represent
$\kappa_{\nu jj'}$ as a Fourier series and restrict the sum up to a
cutoff frequency as
\begin{gather}
\kappa_{\nu jj'}\simeq-\frac{2|\varphi_{\nu j}||\varphi_{\nu
 j'}|}{K_{\nu}N^2}\sum_{k=-N/2+1}^{N/2}|k|e^{\frac{2\pi i}{N}(j-j')k}.
\end{gather}\par
A cluster is built by connecting
sites with the bond probability determined by the dissipative term
$S_0$. In addition, the cluster is flipped with no rejection,
if $\phi_{\nu}{}^{\rm mirror}$ is appropriately chosen so that the potential term
$S_1$ remains unchanged after the cluster is flipped. 
In Fig. \ref{cluster_update}, such rejection-free cluster updates
implemented in this paper are illustrated in
$\phi_\rho$-$\phi_{\sigma}$ planes,
where only the $j$-th path fragment in a cluster is shown.\par
In the upper panel (a), two mirrors are located along $\phi_{\rho}{}^{\rm
mirror}=(n_{\rho}+1/2)\pi$ and $\phi_{\sigma}{}^{\rm
mirror}=(n_{\sigma}+1/2)\pi$,
where $n_{\rho}$ and $n_{\sigma}$ are integers.
Sites are connected with bond probability
\begin{gather}
p_{jj'}={\rm max}\{0,1-e^{-2\sum_{\nu}\kappa_{\nu jj'}s_{\nu j}s_{\nu j'}}\},
\end{gather}
and both the fields $\varphi_{\rho j}$ and $\varphi_{\sigma j}$
in the cluster are sequentially reflected
with respect to the two mirrors, i.e., $(\varphi_{\rho j},\varphi_{\sigma
j})\rightarrow(-\varphi_{\rho j},\varphi_{\sigma
j})\rightarrow(-\varphi_{\rho j},-\varphi_{\sigma j})$. Note that the
connected spins $s_{\nu j}$ and $s_{\nu j'}$ in one channel are not
necessarily parallel, which is different from the original Swendsen-Wang
algorithm.
After this {\it double-field} cluster update,
kink structures connecting nearest-neighbor potential minima are
inserted efficiently.\par
Another cluster update is illustrated in the lower panel (b)
in Fig. \ref{cluster_update},
where only a mirror for the charge degree of freedom is located at $\phi_{\rho}{}^{\rm
mirror}=n_{\rho}\pi$ with an integer $n_{\rho}$.
In this case, a cluster is constructed with bond probability
\begin{gather}
p_{\rho jj'}={\rm max}\{0,1-e^{-2\kappa_{\rho jj'}s_{\rho j}s_{\rho
 j'}}\},
\end{gather}
and the relative fields $\varphi_{\rho j}$ in the cluster are reflected
with respect to the mirror, i.e.,
$(\varphi_{\rho j},\varphi_{\sigma j})\rightarrow(-\varphi_{\rho j},\varphi_{\sigma j})$.
A similar cluster update can be performed also for the spin
degree of freedom.
These {\it single-field} cluster updates insert kink structures between
next-nearest-neighbor potential minima.


Using the PIMC method described above,
we can efficiently simulate the impurity problem in a spinful TLL.
In this paper, we observe
zero-bias conductances of charge and spin channels to directly study the transport
phenomena at low temperatures.
In the linear response regime, a dc conductance at finite
temperature is obtained from analytic continuation
\begin{gather}
G_{\nu}=\lim_{i\omega_n\rightarrow 0}G_{\nu}(i\omega_n),\label{dc_cond}
\end{gather}
where the conductance at a Matsubara frequency
can be calculated from a
correlation function as
\begin{gather}
G_{\nu}(i\omega_n)=\frac{2e^2}{h}\frac{|\omega_n|}{\pi}\int_0^{\beta}{\rm
 d}\tau\langle\phi_{\nu}(\tau)\phi_{\nu}(0)\rangle e^{i\omega_n\tau}.
\end{gather}
Measuring the temperature dependence of the dc conductances (\ref{dc_cond})
for different sets of $K_{\rho}$ and $K_{\sigma}$ near the phase transition,
we determine the phase boundaries in the intermediate region of the
impurity strength $v$.
\section{Results and discussion}

This section is devoted to presenting our PIMC results and the phase diagram
deduced from our data.
We first present and analyze our conductance curves for several given points 
on the $(K_\rho,K_\sigma)$-plane, and determine what kind of phase 
those points belong to.
We then discuss the whole phase diagram, 
and in particular the form and the position of our phase boundaries 
in comparison to their counterparts in the standard perturbative RG picture 
(available only in the weak and strong backscattering limits).
In order to uncover the nature of our phase boundaries, we also attempt 
to give further interpretations to them in the context of quantum Brownian 
motion.

\subsection{Charge and spin conductances and their ``flow'' at low
temperatures}

\begin{figure}[t]
\begin{center}
\includegraphics[width=.8\figurewidth]{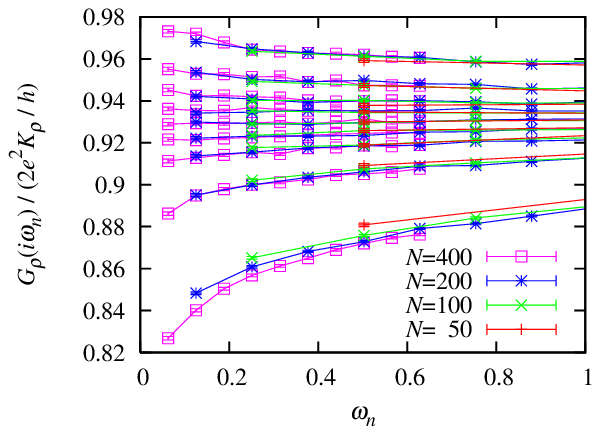}
\includegraphics[width=.8\figurewidth]{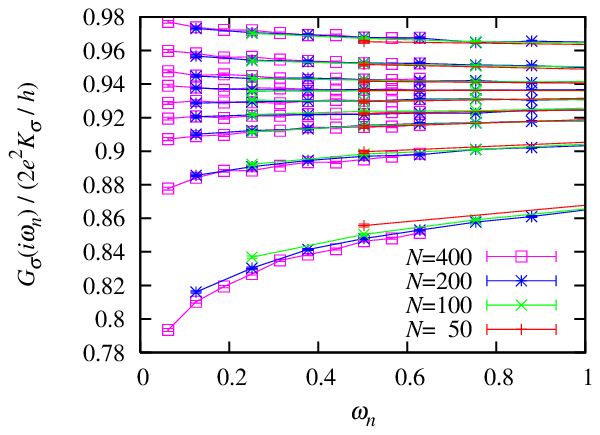}
\caption{\label{GOmegaKs1}(color online)
Symmetric coupling case : $K_\rho \simeq K_\sigma$.
Charge conductance $G_{\rho}(i\omega_n)$ (upper) and
spin conductance $G_{\sigma}(i\omega_n)$ (lower)
for different values of $K_\rho$,
plotted as a function of $\omega_n$.
$K_{\sigma}$ is fixed at $K_{\sigma}=1.0$
$N=$ 50, 100, 200, and 400, and only the first ten points are
shown for each Trotter number $N$.
From top to bottom, the values of $K_{\rho}$ are
1.2, 1.1, 1.05, 1.025, 1.0, 0.975, 0.95, 0.9, and 0.8.
Conductance curves (of both charge and spin)
show an upward bend with decreasing $\omega_n$
when $K_{\rho}>1.025$, whereas they are bent downwrad when
 $K_{\rho}<0.975$.
}
\end{center}
\end{figure}
\begin{figure}[t]
\begin{center}
\includegraphics[width=.8\figurewidth]{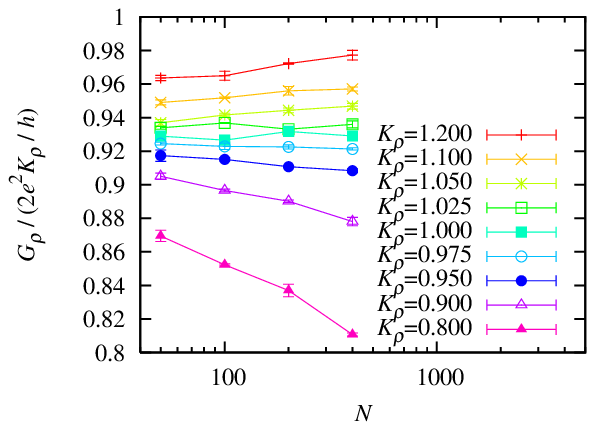}
\includegraphics[width=.8\figurewidth]{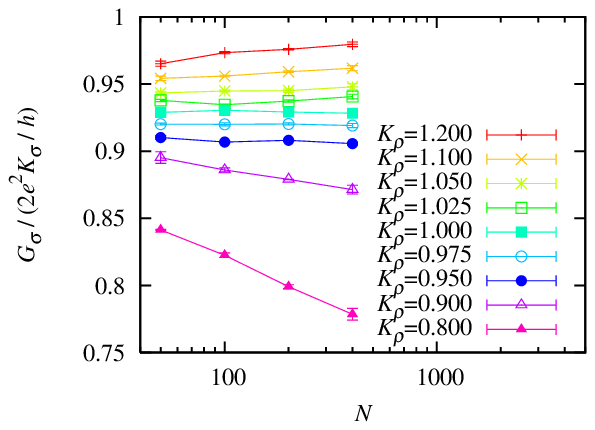}
\caption{\label{GTKs1}(color online) DC conductances deduced from Fig. 3.
Charge (upper) and spin (lower) conductances for different values of $K_\rho$
are plotted as a function of inverse temperature $N$.
$K_{\sigma}$ is fixed at $K_\sigma=1.0$.
Symmetric coupling case ($K_\rho\simeq K_\sigma$).
As expected from Fig. 3,
both charge and spin conductances increase with decreasing temperature
 (increasing $N$)
for $K_{\rho}>1.025$, whereas they decrease for $K_{\rho}<0.975$.}
\end{center}
\end{figure}\

\begin{figure}[t]
\begin{center}
\includegraphics[width=.8\figurewidth]{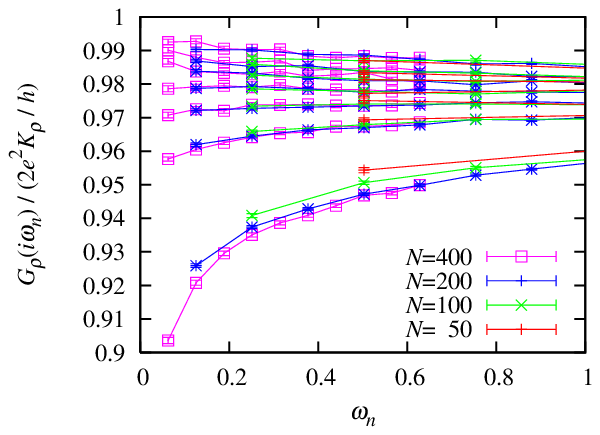}
\includegraphics[width=.8\figurewidth]{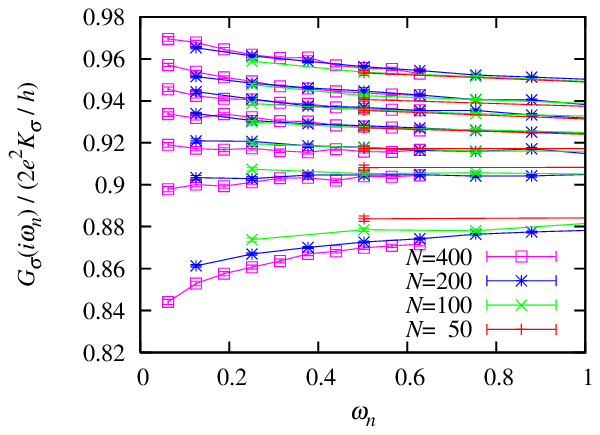}
\caption{\label{GOmegaKs1.8}(color online)
Asymmetric coupling case : $K_\rho\ll K_\sigma$.
Charge conductance $G_{\rho}(i\omega_n)$ (upper) and
spin conductance $G_{\sigma}(i\omega_n)$ (lower)
for different values of $K_\rho$,
plotted as a function of $\omega_n$.
$K_{\sigma}$ is fixed at $K_{\sigma}=1.8$.
$N=$ 50, 100, 200, and 400, and only the first ten points are
shown for each Trotter number $N$.
Charge and spin channels behave quite differently in this parameter regime.
Upper :  $G_{\rho}(i\omega_n)$ is plotted for (from top to bottom)
$K_{\rho}=$ 0.600, 0.550, 0.525, 0.500, 0.475, 0.450 and 0.400. 
With decreasing $\omega_n$,
the charge conductance shows either a {\it monotonous} upward 
(first three curves) or downward (last three curves) bend.
Such a behavior resembles the symmetric coupling case (see Fig. 3).
Lower :
$G_{\sigma}(i\omega_n)$ is plotted for (from top to bottom)
$K_\rho =$ 0.500, 0.450, 0.425, 0.400, 0.375, 0.350 and 0.300. 
The spin channel shows a {\it non-monotonous} behavior
when $K_{\rho}=$ 0.400, 0.375 and 0.350. 
For details, see also Fig. 6.
}
\end{center}
\end{figure}
\begin{figure}[t]
\begin{center}
\includegraphics[width=.8\figurewidth]{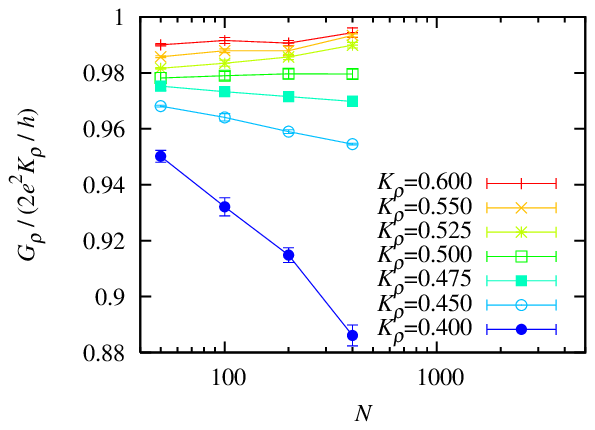}
\includegraphics[width=.8\figurewidth]{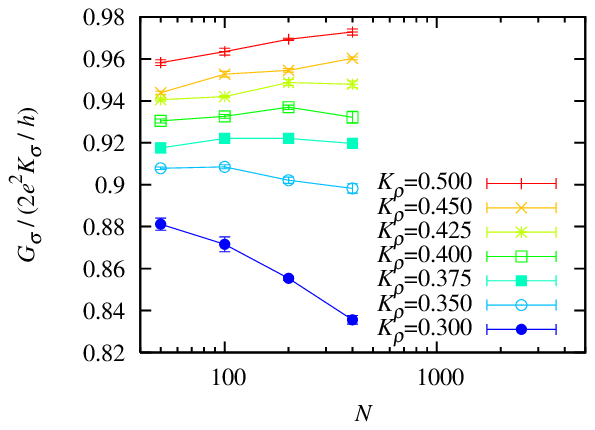}
\caption{\label{GTKs1.8}(color online)
DC conductances deduced from Fig. 5.
Charge (upper) and spin (lower) conductances for different values of $K_\rho$
are plotted as a function of inverse temperature $N$.
$K_{\sigma}$ is fixed at $K_\sigma=1.8$.
Asymmetric coupling case ($K_\rho\ll K_\sigma$).
Upper :
The temperature dependence for the charge channel shows
monotonous increase ($K_{\rho}>0.5$) or decrease ($K_{\rho}<0.5$)
with decreasing temperature.
Lower :
The spin conductance behaves non-monotonously for $K_{\rho}=$ 0.35,
 0.375, and 0.4,
e.g., for $K_\rho=0.375$, the conductance increases from $N=50$ to
 $N=100$, whereas it
decreases from $N=200$ toward $N=400$.
}
\end{center}
\end{figure}

We show in this subsection our simulation results for
$v=4,{\mit{\Delta}}\tau=0.25$, and $N=$ 50, 100, 200, and 400. Here $N$ is
inversely
proportional to temperature as $N=\beta/{\mit\Delta\tau}$.

{\it Symmetric coupling case : $K_{\rho}\simeq K_{\sigma}$} ---
Let us first investigate a domain of ($K_{\rho}, K_{\sigma}$)
for which the phase diagrams in the limit of weak and strong scattering
barriers are smoothly connected. 
This happens when two coupling constants are symmetric, or isotropic
(see Fig. 1). 
In Fig. \ref{GOmegaKs1}, we focus on the $K_{\sigma}=1$ line
(on which the spin part is SU(2) symmetric) and plot the first ten points of 
$G_{\nu}(i\omega_n)$ as a function of $\omega_n$ for different values of
$K_{\rho}$ near the phase boundary. 
For a given $K_{\rho}$,
results for different $N$, i.e., for different temperatures
are superposed to form a bundle of curves.
For both $G_{\rho}(i\omega_n)$ and $G_{\sigma}(i\omega)$,
one can see that the curves for $K_{\rho}>1.025$ are bent upward 
with decreasing $\omega_n$ (in the limit of $\omega_n\rightarrow 0$), 
while the curves for $K_{\rho}<0.975$ are bent downward.
Note also that for a given $K_{\rho}$, the slope of different curves
composing the same bundle always becomes steeper with
decreasing temperature (increasing $N$). 
When the data shows such a monotonous dependence on temperature,
one can determine the phase boundaries by simply identifying a
turning point at which the bend of $G_{\nu}(i\omega_n)$ changes from
upward to downward with decreasing $\omega_n$.
We will see later, however, that the temperature dependence of 
$G_{\nu}(i\omega_n)$ curves can become non-monotonous in the presence of a
non-trivial fixed point.
We have actually determined our phase boundaries by tracing
the temperature dependence of dc
conductances obtained from (\ref{dc_cond}) following
Refs \onlinecite{werner1,werner2}.
In Fig. \ref{GTKs1}, we plot the dc conductance of charge and spin 
as a function of the inverse temperature $N$, which are obtained by
extrapolating the first five points on each curve in
Fig. \ref{GOmegaKs1} to $\omega_n\rightarrow 0$.
With decreasing temperature, the conductances 
shows a monotonous increase (decrease) when $K_{\rho}>1.025$
($K_{\rho}<0.975$). 
Here, the charge and spin channels show a simultaneous transition from conducting 
to insulating phase in consistent with the RG results. 
Recall that in the RG picture (see Fig. \ref{phase_rg})
the IV and I phases touch at $(K_{\rho},K_{\sigma})=(1,1)$
both in the weak- and strong-backscattering regimes,
which suggests that the phase boundary at that point is a straight line
independent of $v$ in the $(K_{\rho},K_{\sigma},v)$-space.\par

{\it Asymmetric coupling case : $K_{\rho}\ll K_{\sigma}$} ---
Let us turn to a parameter regime in which non-trivial RG flow
is expected for a finite backscattering strength.
Such a behavior is actually expected
whenever the phase boundaries in the weak- and strong-impurity limits
are not identical,
but occurs typically when two coupling constants are highly asymmetric,
or anisotropic : $K_{\rho}\ll K_{\sigma}$. 
In Fig. \ref{GOmegaKs1.8}, 
we plot $G_{\rho}(i\omega_n)$ and $G_{\sigma}(i\omega_n)$
with $K_\sigma$ fixed at $K_{\sigma}=1.8$, and for various values of $K_\rho$.
The corresponding temperature dependence of the dc conductances 
is shown in Fig. \ref{GTKs1.8}. 
A careful reader might immediately notice
that the charge and spin channels behave differently
in these two figures.
Of course, the origin of the difference is the anisotropy between $K_{\rho}$ and $K_{\sigma}$,
but let us look into more carefully how they are different. 
In Fig. \ref{GOmegaKs1.8}, 
conductance curves for the charge channel, i.e.,
$G_{\rho}(i\omega_n)$ with $K_\rho=0.600,0.550$ and 0.525 are bent upward
with decreasing $\omega_n$,
while the same curves for $K_{\rho}=0.475, 0.450$ and 0.400 are clearly bent downward.
The temperature dependence of the dc conductance (Fig. \ref{GTKs1.8})
shows a monotonous behavior similar to the case of $K_{\sigma}=1$
(see Fig. 4). 

On the other hand, the temperature dependence of
spin conductance $G_{\sigma}(i\omega_n)$ is more peculiar:
for example, if one focuses on the conductance curves for $K_{\rho}=0.35$,
their slopes are upward at high temperatures, e.g., between $N=$50 and 100, 
whereas the same curves have an opposite slope at low temperatures, e.g.,
between $N=$ 200 and 400. 
Such non-monotonous behaviors might be more clearly seen, if we look into  
the dc conductance in Fig. \ref{GTKs1.8} for $K_{\rho}=$ 0.350,
0.375 and 0.400.

Similar crossover behaviors are observed
whenever analyzing the boundary between
phases I and III, and are also reported in the Josephson junction system studied in
Ref. \onlinecite{werner2}. 
From the RG viewpoint,\cite{kane2}
the unusual temperature dependence of the spin channel 
derives from non-monotonous flows
of $v$ near the intermediate unstable fixed point.
Since the precise location of such a non-trivial fixed point is unknown,
one cannot immediately conclude that the spin channel is
in the conducting phase,
even if the conductance, e.g. for $K_{\rho}=0.450$ or $0.500$,
tends to increase monotonously toward low temperatures
up to $N=400$.
Possibly, it might turn insulating at a certain lower temperature 
which is numerically inaccessible. 
Since it is difficult to locate true phase boundaries at $T=0$ 
in the presence of intermediate unstable fixed points, 
we instead identify the phase
boundary at $T\ne 0$ by observing the temperature dependence near the
lowest temperature $N\simeq 400$.

\subsection{The phase diagram in the
$\bm(\bm K_{\bm\rho}\bm,\bm K_{\bm\sigma}\bm)$-plane for
an intermediate scattering strength}

\begin{figure}[t]
\begin{center}
\includegraphics[width=.9\figurewidth]{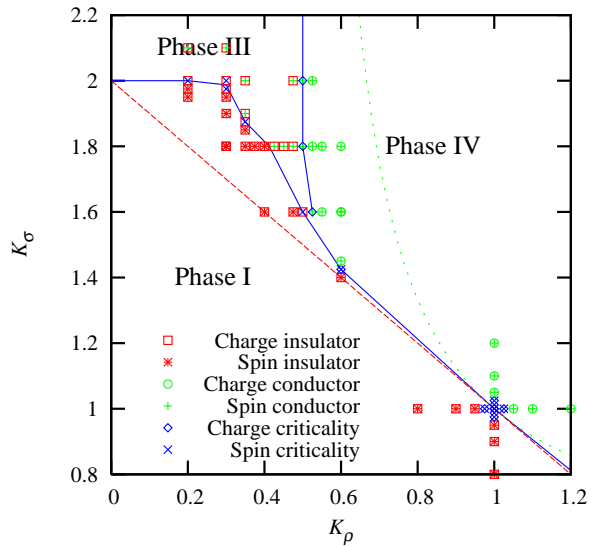}
\caption{\label{phase_pimc}(color online) Phase diagram obtained from
our PIMC simulations for the impurity strength $v=4$.
The solid lines stands for the boundaries between the I, III, and IV
phases.
In order to ease comaprison with the RG results,
we superposed, as a guide to the eyes, the WBS results (dashed straight line) 
for the I-III and I-IV boundaries,
and the SBS results (dotted hyperbolic curve) for the I-IV and III-IV boundaries,
on top of our phase diagram.
}
\end{center}
\end{figure}

Repeating the analyses outlined
in the previous subsection for different sets of 
$K_{\rho}$ and $K_{\sigma}$ near the phase transition, 
we determine the whole phase boundaries in the
$(K_{\rho},K_{\sigma})$-plane. 
In Fig. \ref{phase_pimc}, we show our phase diagram for a finite
impurity strength
$v=4$ obtained from the PIMC 
simulations at inverse temperature
$\beta=400{\mit\Delta}\tau$. 
Due to the symmetry of the action (\ref{act}) in terms of $\rho$ and $\sigma$, 
the following discussion also holds true when
the charge and spin degrees of freedom are interchanged. 
In that case, the phases II and III are, of course, exchanged. 

In Fig. \ref{phase_pimc} one can see three different phases:
(Phase I) neither charge nor spin is conducting,
(Phase III) only spin is conducting,
(Phase IV) both charge and spin are conducting.
The phase boundaries are shown by solid lines.
If one compares them with the phase boundaries obtained
by the renormalization group (RG) analyses
in the weak and strong backscattering limits,
one can verify that all the boundaries are indeed located between the two
limiting cases.
However, the way they are shifted from either of the limits
is not uniform, i.e.,
for nearly isotropic interactions $K_{\rho}\simeq K_{\sigma}$, 
our phase boundary at an intermediate coupling is much closer
to the weak backscattering (WBS) phase boundary,
whereas for strongly anisotropic interactions $K_{\rho}\ll K_{\sigma}$,
our phase boundary between I and III phases is much closer
to the strong backscattering (SBS) phase boundary.
Similarly, our III-IV boundary is much closer to the WBS boundary.
Correspondingly, the tricritical point, i.e., the meeting point of I, III, and IV
phases, lies between its counterparts
in the WBS and SBS limits.\par

For nearly isotropic interactions $K_{\rho}\simeq K_{\sigma}$, 
the obtained critical line for a finite $v$ between insulating (I) and conducting (IV)
phases looks much similar to the WBS boundary:
$K_{\rho}+K_{\sigma}=2$.
This can be understood as a result of the large
conductances of the charge and spin channels (see, e.g., Fig. \ref{GTKs1}).
Then, we can judge that our scattering potential $v$ is relatively small.
For strongly anisotropic interactions $K_{\rho}\ll K_{\sigma}$, 
the III-IV boundary also resembles its WBS counterpart: $K_{\rho}=1/2$,
while the I-III boundary does not.
Interestingly enough, for a broad range of $K_\rho$, say, $0<K_\rho<0.3$, 
the latter boundary is almost superposed on the SBS phase boundary: $K_{\sigma}=2$,
which contradicts a naive expectation from the presumably small
scattering potential.
Thus increasing anisotropy,
the position of our phase boundary shifts from that of WBS to SBS
counterparts.\par

In order to clarify such rather unexpected behavior of the I-III boundary,
here we discuss how vertical and horizontal boundaries appear
in the $(K_{\rho},K_{\sigma})$-plane.
As an example of these boundaries, let us recall the RG phase diagram in
the WBS and SBS limits in Fig. \ref{phase_rg}.
In the WBS case, a vertical boundary appears between III-IV phases,
where the interactions are strongly anisotropic ($K_{\rho}\ll K_{\sigma}$)
and, for $K_{\sigma}>2$, the phase diagram is characterized only by the value
of $K_{\rho}$.
Due to the strong attraction $K_{\sigma}\gg 1$,
the spin channel transmits through the impurity so freely
that the weak scattering potential hardly influences this spin channel.
In that sense, the spin mode is irrelevant
and the transport of the system depends only on the charge mode,
which we call a {\it one-field} situation.
In the SBS case, on the other hand, a horizontal boundary appears
between I-III phases,
where again the interactions are fully anisotropic and, for $K_{\sigma}<1/2$, 
the phase diagram is characterized only by the value of $K_{\sigma}$.
Due to the strong repulsion $K_{\rho}\ll 1$,
the charge channel scarcely go over the impurity,
and is almost extinct.
Then again, we see another {\it one-field} situation.
In both the WBS and SBS cases,
the vertical or horizontal boundary appears as a result of the strong
anisotropy in the interactions, rather than the extreme values of $v$.\par

We can now interpret the unexpected change
 of the I-III phase boundary
observed in Fig. \ref{phase_pimc}.
The horizontal region in the I-III boundary derives from the occurrence
of a {\it one-field} situation where transport of the system is
characterized only by the spin channel.
Thus, after integrating out the extinct charge channel,
one can argue that the effective action is given by
\begin{gather}
S\simeq\sum_{\omega_n}\frac{|\omega_n|}{2\pi
K_{\sigma}\beta}|\tilde{\phi}_{\sigma}(\omega_n)|^2+v\int{\rm d}\tau\cos\phi_{\sigma}(\tau),
\label{1field}
\end{gather}
which takes the same form as the action of a single impurity problem 
in a \textit{spinless} TLL with interaction parameter $K=K_{\sigma}/2$.\cite{kane1} 
Our phase diagram implies that the I-III boundary in the SBS limit: $K_{\sigma}=2$,
partially preserves its position for a broad range of $v>0$,
leading to the robustness of the boundary.
The I-III boundary in Fig. \ref{phase_pimc} also shows
a small deviation from $K_{\sigma}=2$ with increasing the value of $K_\rho$.
In this parameter region, the pinning of charge degree of freedom is
no longer complete, and we believe that the crossover from the one-field
model (\ref{1field}) to the original two-field model (\ref{act}) occurs.
We will give further discussion on this behavior in the context
of quantum Brownian motion in the next subsection.
Note that, due to the dualily of the impurity problem in a TLL,
\cite{kane2,furusaki}
the III-IV boundary will also show a similar crossover for a large value
of $v$.

It should also be added that a model
qualitatively equivalent to the
single impurity problem in a spinful TLL 
is realized in 
a completely different context.
Werner \textit{et al}. have performed the PIMC simulations of a
system with two Josephson junctions, and shown a phase diagram
similar to Fig. \ref{phase_pimc} consisting of three distinct
phases.\cite{werner2} 
Although the kinetic term derived from charging energy $E_C$, which
is absent in our model, does not change the essential nature of the
phase boundaries, it nevertheless influences transport phenomena at
low temperatures. 
In Ref. \onlinecite{werner2}, the authors seem to assume that the system undergoes transition to a one-field model like (\ref{1field})
as soon as a channel enters an insulating phase,
and that the position of the tricritical point
is independent of the Josephson coupling strength $E_J$,
which corresponds to the backscattering strength $v$ in our system.
As we have seen  above, however,
the original two-field model slowly {\it crossovers} to a one-field one,
and so the tricritical point should in general depend on
$E_J$, or $v$. If our conjecture holds true also
for the two-Josephson-junction system,
the discrepancy in the tricritical point discussed
in Ref. \onlinecite{werner2} would be resolved.

\subsection{Interpretation in the context of quantum Brownian motion}

\begin{figure}[t]
\begin{center}
\includegraphics[width=.5\figurewidth]{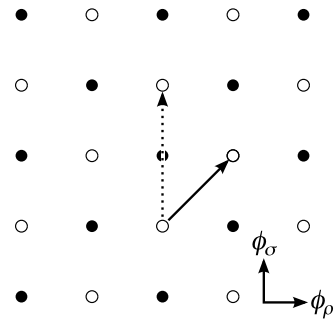}
\caption{\label{tunnelings}
Minima and maxima of the two-dimensional potential
$v\cos\phi_{\rho}\cos\phi_{\sigma}$.
The empty (filled) circles represent the potential minima (maxima).
The solid (dotted) arrow stands for an example of tunneling of a massles
 particle
between minima through a suddle point (over a maximum).
For strongly anisotropic friction,
e.g., $K_{\rho}{}\!\!^{-1}\gg K_{\sigma}{}\!\!^{-1}$,
the diagonal process (through a suddle point) is suppressed so that
the relevant process becomes the vertical one (over a maximum).}
\end{center}
\end{figure}

In the previous subsections, we have seen the system not only undergoes
transition between insulation and conduction for charge and spin channels, 
but also crossovers from the original two-field (\ref{act}) to a
one-field model like (\ref{1field}). 
In order to interpret the crossover more clearly, 
let us reconsider our previous results from the viewpoint of
quantum Brownian motion. 
As is clear from the Caldeira-Leggett form of the action (\ref{act}),
our spinful single barrier problem is equivalent to two-dimensional
dynamics of a massless quantum Brownian particle in a periodic potential. 
In this picture, the bosonic fields $(\phi_{\rho},\phi_{\sigma})$ play the role
of particle's coordinates.
As shown in Fig. \ref{tunnelings}, the potential
$v\cos\phi_{\rho}\cos\phi_{\sigma}$,
for $v>0$, has minima (maxima) at $(n_{\rho}\pi,n_{\sigma}\pi)$
with integral $n_{\rho}$ and $n_{\sigma}$ such that
$n_{\rho}+n_{\sigma}={\rm odd\,(even)}$.
Each minimum corresponds to a certain ground state where integral numbers of 
electronic charges and spins exist in the $x>0$ part of the system. 
The dissipation strengths in the
$\phi_{\rho}\,(\phi_{\sigma})$ direction is proportional
to $K_{\rho}{}\!\!^{-1}\,(K_{\sigma}{}\!\!^{-1})$.\par

For nearly isotropic
dissipations $K_{\rho}{}\!\!^{-1}\simeq K_{\sigma}{}\!\!^{-1}$,
the particle at low temperatures tunnels from one minimum to another,
usually through a saddle point between them, and only occasionally over
a maximum.
Although charge and spin are generally separated and propagate with
different velocities in a TLL, tunneling through a saddle point
recovers the spin-charge combined nature of a physical electron. 
On the other hand,
for strongly anisotropic dissipations, e.g., $K_{\rho}{}\!\!^{-1}\gg K_{\sigma}^{-1}$,
the position of the most relevant tunneling process
could be taken by the other
If the friction in the $\phi_{\rho}$ direction is large enough to suppress
completely
the tunneling through a saddle point,
the particle can only go over a potential maximum
in the $\phi_{\sigma}$ direction toward another minimum.
In this case, the system is dominated by the one-dimensional action
(\ref{1field})
describing the horizontal part of I-III boundary in Fig. \ref{phase_pimc}.
If the friction in
the $\phi_{\sigma}$ direction is so small that the massless particle can move
freely in that direction, the system depends only on the $\phi_{\rho}$
coordinate and one gets another one-dimensional
action analogous to (\ref{1field}).\par

If we now return to the original TLL picture, the crossover from 
a two-field to a one-field model depends on the anisotropy of interactions
and the impurity strength $v$ in a complicated way. 
Moreover, the phase diagram in Fig. \ref{phase_pimc} shows
a crossover behavior 
in the $K_{\rho}$-$K_{\sigma}$ plane for an intermediate anisotropy, 
where the effective action can no longer be written in such a simple form as
(\ref{1field}). 
It is worth noting that in the small- and large-barrier limits,
crossovers to a one-field model occurs in so small a region
that one cannot observe them in the phase diagram in Fig. \ref{phase_rg},
while the PIMC simulation does
demonstrate that they could actually appear in a broad range of $v$.

\section{summary}
In this paper, we have studied the single impurity problem in a spinful TLL
using the path-integral Monte Carlo methods.
Measuring the temperature dependence of the charge and spin conductances,
we have obtained the phase diagram characterized by perfect conduction
or insulation of the charge and spin channels,
which is consistent with the renormalization group (RG) results
in the weak- and strong-impurity limits.
We have also observed non-monotonous temperature dependence of conductances,
which qualitatively supports the non-linear flows
near non-trivial unstable fixed points predicted in the RG picture.
The phase diagram obtained from our simulations for an impurity
with intermediate strength shows unexpected shift of a phase boundary
for strongly anisotropic interactions.
By mapping the impurity problem to a quantum Brownian motion picture,
we have proposed an intuitive interpretation of this behavior
from the viewpoint of crossover to a one-field model.
\begin{acknowledgments}
Y.H. is grateful to T. Matsuo for teaching him the PIMC methods.
We also thank K. Kamide for stimulating discussions.
The computation in this work has been done using the facilities of the
Supercomputer Center, Institute for Solid State Physics, University of Tokyo.
\end{acknowledgments}
\newpage
\bibliography{thesis1.bib}
\end{document}